**Challenge-Based Funding to Spark Origins Breakthroughs**

Cole Mathis[1,2] and Harrison B. Smith[3,4]

[1] Biodesign Institute, Arizona State University, Tempe AZ, USA
[2] School of Complex Adaptive Systems, Arizona State University, Tempe AZ, USA
[3] Earth-Life Science Institute, Institute of Science Tokyo, Tokyo, Japan
[4] Blue Marble Space Institute of Science, Seattle WA, USA

At some point in their career, astrobiologists find themselves sitting on a plane next to a curious traveler who, after hearing that you're studying how life came to be, inevitably asks: *so what have you all learned in the past few years?* This is a difficult question to answer. Has our understanding of the origins of life improved in the last decade? On the one hand, a considerable number of results have been published across a variety of disciplines, each aiming to advance our understanding of the origin of life [1,2]. On the other hand, the field's biggest trophies seem as unobtainable as they have always been [3–5]. We believe NASA should coordinate the identification of key measures of success in origins research, and deploy novel funding mechanisms to incentivize the pursuit of those benchmarks.

Origins of life research is marred by the problem of ambiguous or open ended questions and goals. Because of so many unknowns and disagreements about definitions and theories, it is not, and will not, be clear how to know when such goals have been achieved. For example, one of the goals of the origins of life community is to make "life in the lab"—a formidable task, to put it lightly—but perhaps even more difficult will be to convince others that this was achieved. This is because "life in the lab" is not yet a well defined concept, because *life* is not yet a well defined concept. These kinds of indefinite goals do not lend themselves to incremental progress because they deemphasize advancements which could be made through well-defined objectives .

In contrast to other astrobiology subfields in which memorable advancements are made on the cadence of major missions, progress in origins of life research comes foremost from the lab. The absence of megaprojects means that the community is never required to focus its energy, or reconcile its differences—yielding ambiguous and disjoint experimental results without the weight to coalesce into substantive breakthroughs. As a way to overcome this, we argue that the origins community should be focused on goals that have agreed upon meaning, and can be consensually categorized as achieved or unachieved. The origins community needs these goals to maintain coherence amongst a federation of problems with the shared, but nebulous aspiration of understanding the origins of life.

Understanding the origin of life is nebulous at present because *life* as we understand it is not an isolated process but a constellation of co-occurring phenomena. To overcome this obstacle, we propose drawing inspiration from the artificial intelligence (AI) community. Intelligence is difficult to define for many of the same reasons life is difficult to define. In the face of this conceptual challenge, segments of the AI community focused on achieving interesting outcomes; even if those outcomes did not make intelligence itself. For example, while Large Language Models are not considered intelligent, they are instructive for demarcating language usage from the broader phenomena of intelligence by achieving certain benchmarks

---

[6,7]. The origins of life community should follow suit, by identifying goals and quantitative benchmarks that measure progress towards them. Even if the generated phenomena are not unequivocally life-like, demonstrating systems that achieve these goals will sharpen the distinction between life itself and the constellation of phenomena that co-occur with life.

We propose a list of challenges with clear *Finish Lines*, similar to the X-prize, (and to a lesser extent the Hilbert problems), as a jumping off point for the origins of life community [8,9]. Each Finish Line is a description of what will be achieved if the goal is reached. The intent is not to impose, top down, what should be areas of inquiry, but instead to compel the community to coalesce around explicit problems of the highest priority, in the way that the physics, astronomy, and planetary science communities are forced to do when setting science objectives for missions and megaprojects [10,11]. We limited this list to activities which can in principle be achieved on Earth, because while many tasks involving planetary exploration would yield considerable insight into the origins of life, these tasks would ultimately be within the purview of large scale missions which have their own objectives. The Finish Lines outlined here are in no way exhaustive, but are designed to stimulate research that will ultimately provide evidence for or against the continuation of certain research programs. Ideally reaching (or failing to reach) these Finish Lines will prune active lines of inquiry so that new ideas can be explored.

## Finish Line: Co-maintaining the reaction transformations at the center of biological sugar metabolism without contemporary biochemistry

**Why it's important:** The pentose phosphate pathway (PPP) and its reverse, the Calvin-Benson cycle, are some of biology's most ubiquitous and elegant examples of metabolic engineering [12,13]. These pathways use five key transformations—aldolase, aldose-ketose isomerase, phosphohydrolase, transaldolase, and transketolase—to perform sugar interconversions [12]. These allow cells to convert between different sugar phosphates with extraordinary flexibility, effectively creating a "sugar algebra" system that can generate precisely the carbohydrate molecules needed for various cellular processes [14]. This includes producing essential NADPH for biosynthesis and oxidative stress defense, ribose-5-phosphate for nucleotides, and erythrose-4-phosphate for aromatic amino acids. The pathway's ability to both generate reducing power and interconvert sugars makes it a cornerstone of cellular metabolism, bridging energy production, biosynthesis, and stress response. This suggests these chemical transformations were fundamental to early life, possibly even predating modern enzymes, as some can occur spontaneously under primitive conditions [13]. The simplicity and perceived ancientness of this system make it an ideal candidate to instantiate abiogenically, providing insight into the stability and robustness of core features of biochemical networks in the absence of surrounding biochemistry and contemporary proteins.

## Finish Life: Producing detectable abundances of high assembly index molecules using one-pot abiotic chemistry

**Why it's important**: Life is hypothesized to be unique in its ability to produce large quantities of molecules of high complexity [15]. Assembly theory has emerged as a technique to quantify the complexity of molecules–algorithmically and via measurements of chemical samples [16].

Only molecules associated with living systems have only been observed to reach high molecular assembly indices (MA) [15]. The highest indices observed for molecules which are produced abiogenically include species like tryptophan, with an MA of 14, and direct evidence suggest covalently bonded molecules of MA>15 cannot be detected outside living systems [15]. Demonstrating the emergence of detectable quantities of high MA molecules via simple 1-pot experiments, and single-carbon precursors, would provide a robust example of molecular complexity from abiotic conditions. What is the high MA molecule that can be produced in these conditions? To be relevant to the origins of life the conditions that yield these molecules should also yield other complex molecules (e.g., the conditions should not be so fine tuned as to only produce a single compound). Experiments of this type would help bridge the gap between abiotic and biotic chemistry, while shedding light on the relevant abiotic processes which are able to produce complex molecules in the absence of life's complex chemical infrastructure.

**Finish Line: Demonstrating that fundamentally different product distributions can arise from a single initial condition**

**Why it's important:** A common theme in contemporary astrobiology research, and origins of life research is that terran biochemistry represents but one possibility out of many [17–20]. Consequently, many scientists assume that life on different worlds could use fundamentally different biomolecules [21]. It is not clear (i) if this is possible in principle, and (ii) if this could happen even after the planetary boundary conditions are fixed. To address this, attempts should be made to design chemical experiments that could yield profoundly different and varied outcomes from fixed initial conditions. Phenomena of this type can be observed in a variety of artificial life models, but have not been systematically investigated in empirical systems [22]. The challenge here would be to design an experimental platform, likely based on simple molecular building blocks (e.g. a Miller-Urey type system) in which identical (up to measurement accuracy) initial conditions yield qualitatively different product distributions. This could occur through a type of dynamical chaos that is amplified by ratchet effects, locking in certain subsets of chemical possibilities. The observable consequence of this would be a system that reliably produces outcomes that cannot be reproduced—but the property of not being able to be reproduced is itself reproducible. Confirming this characteristic feature would represent a fundamental challenge in analytical chemistry, requiring cross-cutting expertise from both prebiotic chemists and planetary protection experts to limit contamination. If such a system cannot be designed, it would indicate that the boundaries of biochemistry are completely cast by geochemical planetary conditions. Meanwhile if reproducible-irreproducibility in chemistry can be demonstrated it would suggest detailed knowledge of planetary conditions are less important than currently perceived, as they would ultimately be washed out through ratcheting processes in chemical evolution.

**Finish Life: Creating a population of self-replicators that can spontaneously bypass Eigen's error threshold**

**Why it's important:** The error threshold can be expressed generally as the number of mutations that a replicating polymer can sustain per monomer before reliable information transfer becomes impossible [23,24]. If the mutation rate is too high, then the longest polymers which fall below the error threshold are only on the order of ~100 monomers, which are much smaller than the smallest known encoding of error correcting enzymes. While the error threshold could be overcome by the existence of lower mutation rates, higher mutation rates are likely still problematic in a number of scenarios. A number of proposed solutions to the error threshold have been investigated via models, but have as of yet, not been demonstrated experimentally [25–27]. Experimental demonstration of a self-replicating population of polymers, either by implementing an existing proposed model solution, or another solution, would resolve this outstanding paradox in the early evolution of life.

**Finish Line: Evolving a self-replicating RNA system to use DNA**

**Why it's important:** The RNA world "hypothesis," is one of the most commonly cited, and heavily studied areas of origins of life research [3]. At its core, the concept is that earlier living systems may have used RNA as both information and genetic materials. That early life may have used a different set of macromolecules is not unique to the RNA world. But it's not clear how, or under what conditions, a transition between macromolecular paradigms could be facilitated. This conjecture could be explored using synthetic biology experiments in a few different ways. First, directed evolution experiments could attempt to evolve a ribozyme capable of translating RNA catalysts into DNA sequences and (in the same experiments) another ribozyme to translate DNA sequences into RNA catalysts. Alternatively, if it was possible for a selective environment to drive the evolution of DNA utilization, it should be possible to identify environmental conditions which select for RNA genetic information storage [28]. While this finish line may seem secondary to the "holy grail" of RNA research (the identification of an RNA self-replicator), it is actually critical for determining whether RNA based research is relevant to origins of life research at all. If macromolecular systems cannot be transitioned through gradual selective processes, continued development of a RNA world chemistry is irrelevant for understanding the origins of life on Earth (though it may be relevant to modern biology, and alternative biochemistries). This is related to, but distinct from, the challenge of understanding the origin of the genetic code, as the problem of translating between RNA and DNA does not require a code in the way translating between protein and RNA does [29].

**Conclusion:** We believe focusing on intermediate challenges—understandable, and precisely measurable within current paradigms—will best drive progress in the origins field. We have excluded obvious "holy grail" objectives, such as the synthesis of an RNA self replicator, the spontaneous emergence of a genetic code, or the stabilization of a non-enzymatic carbon fixing autocatalytic loop, as we believe these objectives are already implicitly pursued by the field.

The challenges suggested here provide unambiguous targets that should be refined and expanded through community input. This input should not simply recapitulate existing disjoint or overarching research objectives in the community. The Finish Lines here may cut across research paradigms within and between the existing Research Coordination Networks (RCNs).

We suggest that NASA, as the key funding agency, should facilitate the curation of an expanded list of community submitted Finish Lines, and explore innovative new funding schemes to stimulate progress towards their completion. The regulatory and legal challenges in establishing novel funding paradigms are significant, but surmountable. We draw inspiration from the DARPA Cyber AI challenge, which rewarded teams financially for producing a specific outcome [30]. We also draw inspiration from the International Polar Year as an example of the kind of megaproject which would serve the astrobiology community—originally organized in 1882, the idea was to coordinate the collection of meteorological data in the vast polar regions by standardizing and focusing direction between scientists [31]. In the same way, we believe NASA Astrobiology can innovate in this space, by exploring new award types and coordinating the origins community to converge on viable Finish Lines. Funding oriented around Finish Lines could allow researchers to both apply for funding to investigate ideas (as currently implemented), and also reward teams that can demonstrate new technology or milestones.

It's not clear how the origins of life community will arrive at the ultimate achievement: generation of de novo life in the lab. But from the body of knowledge that exists, we can imagine many paths forward. The challenges outlined here aim to focus those pathways, in one way or another, so that we can take some small steps forward. We believe that measurable steps forward are preferable to the paralysis caused by infinitely divergent possibilities.